\documentclass[10pt, conference]{IEEEtran}
\IEEEoverridecommandlockouts
\usepackage{cite}
\usepackage{amsmath,amssymb,amsfonts}
\usepackage{algorithmic}
\usepackage{graphicx}
\usepackage{textcomp}
\usepackage{xcolor}
\usepackage{subcaption}
\usepackage{tabularx}
\usepackage{array}
\usepackage{listings}
\usepackage{epstopdf}
\usepackage{comment}
\usepackage{hyperref}
\usepackage{multirow}
\usepackage{balance}
\usepackage{hhline}
\usepackage{color,soul}
\usepackage{boxedminipage2e}

\graphicspath{{figures/}}

\def\BibTeX{{\rm B\kern-.05em{\sc i\kern-.025em b}\kern-.08em
    T\kern-.1667em\lower.7ex\hbox{E}\kern-.125emX}}

\begin{document}

\title{Checkpointing and Localized Recovery \\for Nested Fork-Join Programs }

\author{\IEEEauthorblockN{Claudia Fohry}
  \IEEEauthorblockA{Research Group Programming Languages /
    Methodologies \\
  University of Kassel, Germany\\
    fohry@uni-kassel.de}
}

\maketitle

\begin{abstract}
While checkpointing is typically combined with a restart of the whole
application, localized recovery permits all but the affected processes
to continue. In task-based cluster programming, for instance, the
application can then be finished on the intact nodes, and the lost
tasks be reassigned.

This extended abstract suggests to adapt a checkpointing and localized
recovery technique that has originally been developed for independent tasks
to nested fork-join programs. We consider a Cilk-like work stealing
scheme with work-first policy in a distributed memory setting, and
describe the required algorithmic changes. The original technique has
checkpointing overheads below 1\% and neglectable costs for recovery,
we expect the new algorithm to achieve a similar performance.
\end{abstract}

\section{Introduction}

Checkpoint/Restart (C/R) is the current standard technique to handle
fail-stop failures of processes in clusters~\cite{exa,ftHerault}. It
is often criticized for limited scalability, and therefore variants such
as uncoordinated~\cite{uncoo}, in-memory~\cite{inMem} and
multi-level~\cite{multiLevel} checkpointing have been devised. These variants reduce the costs, but most
systems still require to restart the whole application when a failure
occurs.

From both a deployment and performance point of view, it may be
preferable to continue the program execution on the reduced set of
processes, which is called \emph{shrinking recovery}. Related to that, a \emph{localized recovery} approach
confines the failure handling to the affected
processes, ideally without any involvement of the others.

A few localized, and partly shrinking, recovery techniques have already been suggested,
notably in the context of task-based parallel
programming~\cite{thibault,krishna,fgcs}. This context is generally
promising for the provision of resilience: Since tasks have clearly
defined interfaces, checkpointing at the task level allows to nicely
combine ease-of-use and efficiency. Thereby ease-of-use is achieved
through a transparent implementation in the runtime system, and efficiency through
saving task descriptors only.

The importance of task-level checkpointing is likely to increase with
the current rise of task-based parallel programming
(e.g.~\cite{chapel,hpx,legion,parsec,tascell,yewpar2,fahringerSurvey,taskbench}).  While the cited
environments differ widely in their mechanisms for task creation and
cooperation, this paper solely refers to nested fork-join programs (NFJs), which were
popularized with Cilk~\cite{cilk5,cilkNOW,satin}.
Listing~\ref{fig:code} depicts the computation of Fibonacci numbers as an example.\\
\vspace*{-0.5cm} \ \\

{\small
  \begin{center}
    \begin{lstlisting}[float=h, caption={Nested fork-join program},label=fig:code,belowskip=-10pt,lineskip=-0.0ex,,]
  int fib(int n) {            // 0
    if (n < 2) return n;
    int x = spawn fib(n-1);   // 1
    int y = spawn fib(n-2);   // 2
    sync;                     // 3
    return x + y;
  }
  \end{lstlisting}
  \end{center}
}

NFJs start with a single root task, here \verb|fib(n)|. Then each task
may spawn any number of children and pass parameters to them. A task
must wait for the results of all children, either explicitly with
\verb|sync|, or implicitly at the end of the function.
We assume that the tasks communicate through parameter passing
and result return only, they must not have side effects.

The execution of a fork-join program gives rise to a tree, such as the
one in Figure~\ref{fig:fib}. In the figure, rectangles denote spawned
functions. Numbers~0 to~3 correspond to sequential code
sections as given by the comments in Listing~\ref{fig:code}. For
instance, section~0 runs from the beginning of the function until
the spawn of the first child.  Downward edges (solid) mark spawns,
and upward edges (dotted) mark result returns at explicit or implicit
\verb|sync|'s.  
NFJ implementations commonly use work-first work stealing, which is explained in
Section~\ref{sect:background}.
\\
\vspace*{-0.4cm} \ \\

\noindent\begin{minipage}{\columnwidth}
\center
\includegraphics[width=0.95\columnwidth]{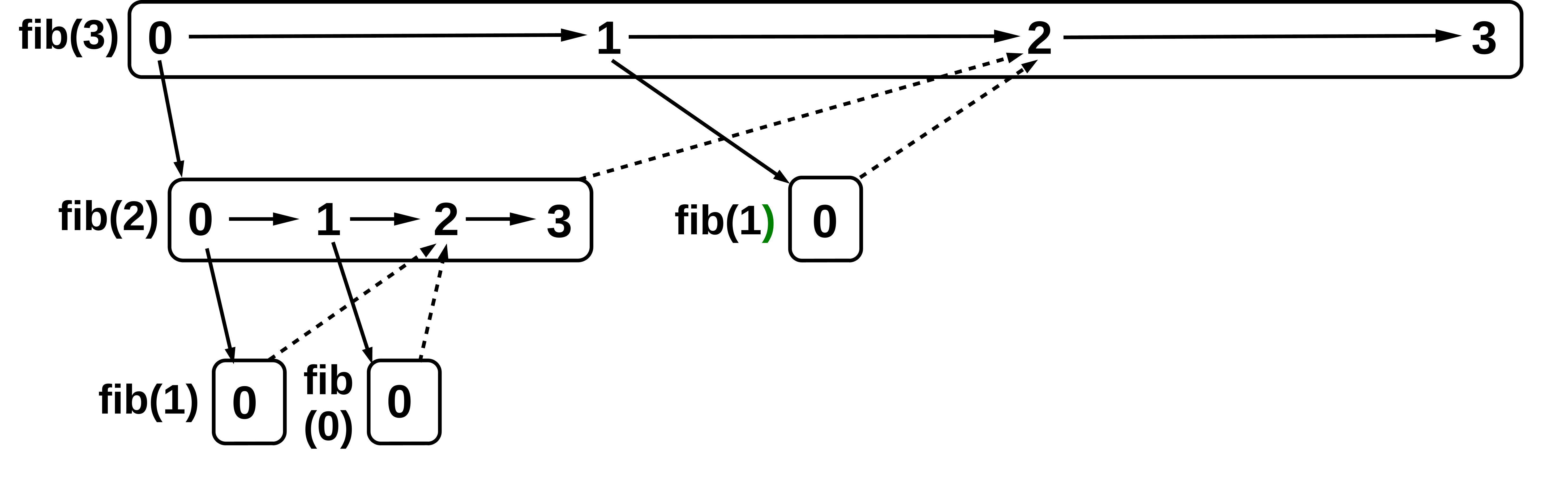}
\captionof{figure}{Execution of nested fork-join programs\\  \ \\}
\label{fig:fib}
\end{minipage}

A resilience scheme with shrinking localized recovery for NFJs under work-first work stealing has already been
introduced by Kestor et al.~\cite{krishna}. Their technique exploits the particular
NFJ style of synchronization to restrict task
re-execution to the lost (sub-)tasks. More specifically, when $k$ out
of $p$ processes fail, a share of \mbox{$k/p$-th} of the previous work must
be re-done on average. The technique has overheads below 1\% in failure-free
runs, but some drawbacks in recovery:
\begin{itemize}
  \item Unlike on average, up to 100\% of the previous work must be
    re-done in worst-case scenarios. These occur when the root of a
    large (sub-)tree fails at the moment when all results have been
    returned there. 
    To avoid such unlimited information loss, the
    authors of~\cite{krishna} suggest to combine their technique with
    standard C/R, which however means to essentially give up localized
    recovery.
  \item When a failure occurs, \emph{all} processes must inspect a
    data structure and participate in a global reduction, 
deviating from perfectly localized recovery.
\item The average share of $k/p$ may still be large for long-running
  programs in failure-prone environments.
\end{itemize}

Therefore, this paper advocates a checkpointing-based alternative.
Checkpointing algorithms already exist for other classes of task-based
parallel programs~\cite{thibault,mcoX10}. We refer to an algorithm for
dynamic independent tasks (DIT) from Posner et al.~\cite{fgcs}, and
adapt it to NFJ. Like NFJ tasks, DIT tasks are spawn dynamically,
forming a tree. However, child tasks do not return a result to their
parent, but contribute to a final result that is accumulated locally
and calculated by reduction.

A recent study has compared the algorithms from Kestor et al.~\cite{krishna}
and Posner et al.~\cite{fgcs} in DIT, to which the NFJ algorithm was
transferred~\cite{unserSubmitted}. The study reported overheads below 1\% for
both algorithms, with those of the NFJ-specific
algorithm~\cite{krishna} being lower in failure-free cases, and
those of the checkpointing algorithm~\cite{fgcs} being lower during
recovery.  We expect similar results for
NFJ, since the algorithm proposed in this paper closely resembles the
original one.

Section~\ref{sect:background} of this abstract states our assumptions
on the failure model and provides background on work stealing. Then
Section ~\ref{sect:newAlgo} sketches the checkpointing algorithm and
explains our proposed algorithmic changes. Section~\ref{sect:related}
outlines some related work, and Section~\ref{sect:conclusions}
finishes with conclusions.

\section{Background and Assumptions}\label{sect:background}

\paragraph{Failure Model} We consider fail-stop failures of processes after permanent hardware failures, and
assume reliable network communication.  Any number of failures may
strike at any time, including simultaneous failures and failures
during recovery. A program must always compute the correct result,
except that it may abort when the resilient store used for checkpoint
saving (see Section~\ref{sect:newAlgo}) fails.  All processes must be
notified of all failures, possibly with a delay.
\\
\vspace*{-0.4cm} \ \\

\noindent\begin{minipage}{\columnwidth}
\center
\includegraphics[width=\columnwidth]{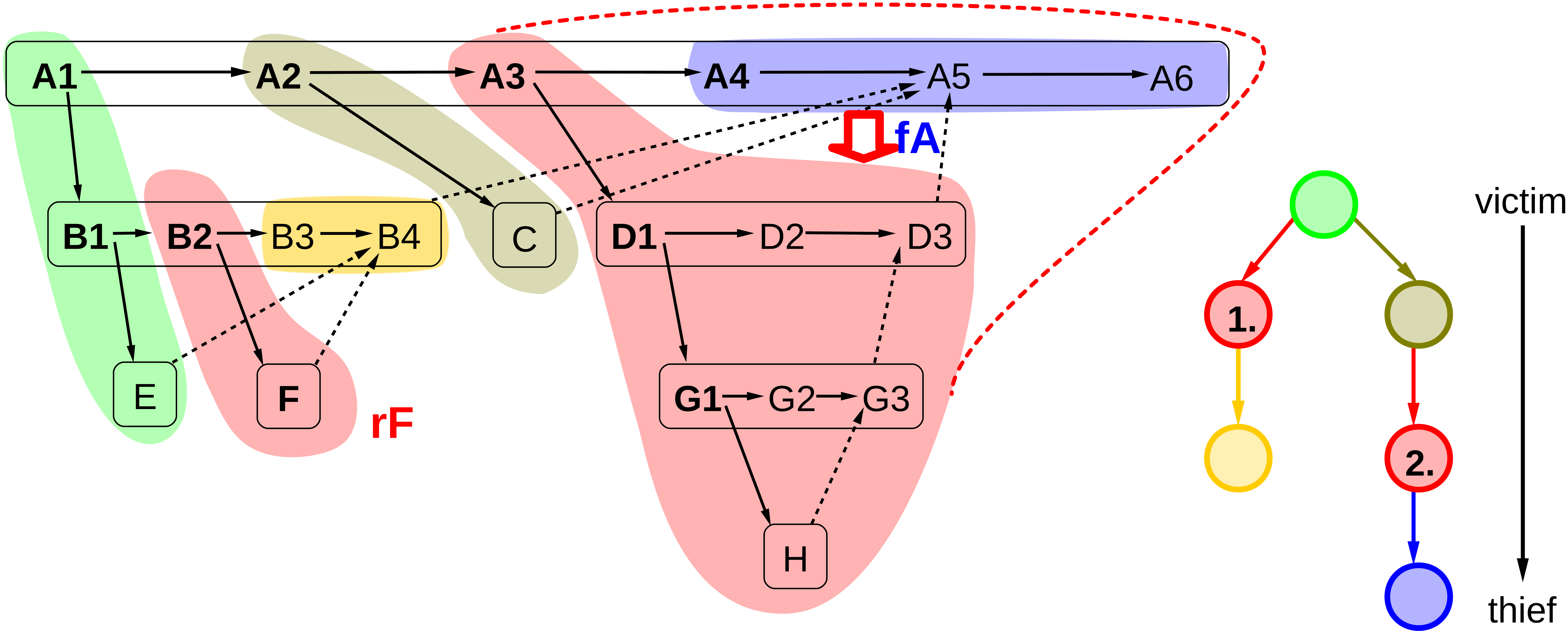}
\captionof{figure}{Work stealing example\\  \ \\}
\label{fig:workFirst}
\end{minipage}

\paragraph{Work stealing} Tasks are commonly executed by a
fixed set of \emph{workers}, which in our distributed memory setting
correspond to processes.  Each worker owns a local pool for storing
and retrieving tasks, which are represented by stack
frames. When the pool is empty, the worker becomes a \emph{thief} and
tries to steal tasks from a \emph{victim}, e.g. from a random
worker. Like~\cite{krishna}, we consider private pools, i.e., the
thief must send a request message for that, and the victim answers by
sending \emph{loot}~\cite{acar,unserCooperative}.

An NFJ execution starts with one local pool holding the root task. Then,
most implementations proceed in a work-first manner: When a
worker encounters a spawn, it branches into the child and puts the continuation of the
parent frame into the pool. Steals always
extract the oldest frame.

Figure~\ref{fig:workFirst} illustrates work-first work stealing. The
figure uses the same notation as Figure~\ref{fig:fib}, but a different
task structure to facilitate further discussion. Each color marks the work performed by a particular
worker.

The computation starts with the green worker (called Green) processing the A frame. At the
first spawn, Green branches into~B, and Brown steals the
continuation of A.  In general, thieves process parent frames, and
victims process children, as shown on the right side of the
picture.

There are only few cluster implementations of the above
scheme~\cite{cilkNOW,krishna}. The one in~\cite{krishna} uses active
messages in a Partitioned Global Address Space (PGAS) setting.  When a thief encounters a \verb|sync|, it
sends the frame back to the victim (or transitively to all victims)
for result matching.  Note that the parent frame is sent back to the
child (and not vice versa), even though the arrows in Figure~\ref{fig:workFirst} indicate that logically the result is
incorporated into the parent frame.
When a victim finishes a task whose parent is
away, it locally saves the result and steals a new task.

For an example, consider Red in
Figure~\ref{fig:workFirst}. It stole frame~B from Green at B2, and was
stolen from by Yellow at B3. Red finished F before Yellow returned. So
Red kept the result (called \verb|rF|) and stole the A frame at
A3. Later, Blue stole the A frame at A4 and already returned it
(called \verb|fA|) at the \verb|sync| opening A5 (as marked by the dotted red line). Now consider the
time when all sections printed in bold have been finished, i.e., right
before branching into~H.  At this time, Red holds \verb|rF|,
\verb|fA|, a local pool with D2
and G2, and a descriptor of~H. Furthermore, Red knows the identities
of all victims and thieves with still unmatched results:
Green and Yellow for the B2 frame, and Brown and Blue for the A3
frame. In its entirety, this information forms Red's
\emph{state}, which will be defined in
Section~\ref{sect:newAlgo}.

\section{Checkpointing Algorithm}\label{sect:newAlgo}

We refer to the AllFT algorithm from Posner et al.~\cite{fgcs},
which is the simplest of three DIT algorithms suggested in that
reference. Our techniques may extend to their incremental scheme,
which reduces the overhead for large task descriptors.

AllFT regularly writes uncoordinated checkpoints to a resilient
store. Reference~\cite{fgcs} uses the IMap of
Hazelcast~\cite{hazelcast} for this purpose, which is a replication-based in-memory
store, but the algorithm is not restricted to it.
Checkpoints for DIT comprise the
local pool contents and the accumulated worker result. They are written
between finishing one task and starting the next one, so that they
capture a coherent state. Beside regular checkpoint writing, AllFT
updates the checkpoints at each steal. A resilient steal protocol ensures
consistency between victim, thief, and their respective checkpoints,
despite possible failures~\cite{fgcs}.

For recovery, each worker has a designated buddy worker. The buddy
reads the last checkpoint of the failed worker from the resilient
store, and inserts the saved tasks into its own pool. This way, all
recently run tasks of the failed worker are re-executed. Moreover,
the buddy takes care of any recently extracted loot from the failed worker.
Only relevant in~\cite{fgcs}, it does not
adopt the accumulated result, which is kept in the resilient store
instead.

Buddies are chosen to be the next worker alive in a ring of workers,
using some numbering of workers.  If a buddy fails, its successor
takes its role. Like stealing, task adoption is protected by a
resilient recovery protocol.

Details of the steal and recovery protocols can be found
in~\cite{fgcs,ijnc}. Briefly stated, they consider specific cases and for
each define a set of actions to be taken to get back to a consistent
state~\cite{fgcs}. For instance, a victim may have to take back tasks
when the thief fails.  Outside the protocols, the checkpointed state
of a worker is always consistent with the ongoing computation.  Thus,
one may safely reset a worker's state to the checkpointed one, without
adjusting the states of the others.

The above algorithm can be adapted to NFJ with only two major changes:
\begin{enumerate}
\item The contents of the checkpoints must be equated to the state of
  an NFJ worker, as defined below.
\item An additional frame return protocol is required.
\end{enumerate}
As in~\cite{fgcs}, checkpoints are written independently for
each worker and contain the worker's state. Since they are written between task processings, they need
not include the internal state of a current task. We define the \emph{state} of a
worker~W to consist of:
\begin{itemize}
\item the current contents of W's local pool,
\item all locally saved task results at W that have
  not yet been incorporated into their parent frame (e.g. \verb|rF|)
  \item all frames returned to W from their thieves that are awaiting result incorporation 
    (e.g. \verb|fA|),
  \item the identities of all victims of~W to which~W has not yet
    returned the respective stolen frame,
    \item the identities of all thieves of~W
      that have not yet returned their frame to~W, and
    \item if relevant, a task descriptor 
      of the next task.
\end{itemize}
The last item is relevant in the following case~1 of possible occasions for
checkpoint writing, but not in case~2. 
When a checkpoint is due, whichever of these two occasions comes first applies:
\begin{enumerate}
\item[1.] right before branching into a child or into a stolen task (e.g. before branching into~H), or
\item[2.]  after finishing a task and incorporating its result into the parent frame or storing it locally (e.g. after finishing~H and incorporating its result into the G frame).
\end{enumerate}
In case~2, the descriptor of the next task is irrelevant, since it
will either be a task from the local pool, which is part of the state
anyway; or a newly stolen task, for which the steal protocol will
schedule another checkpoint.

 The steal and restore protocols
can essentially be taken from~\cite{fgcs,ijnc}, since the handshaking
to reach consistency is independent from the contents of
checkpoints. Merely a few differences exist in the way in which the
buddy worker adopts the failed worker's data. Most importantly, for
stored results like \verb|rF|, it must inform the thief (in the
example: Yellow) about the result's new location.  This is feasible
since the identities of the thieves are contained in the adopted
checkpoint. If the thief has failed as well, the buddy contacts
the buddy of the thief instead, which has adopted or will adopt the
stolen frame or a continuation thereof.
The identity of the thief's buddy can be
figured out easily, since it is the next worker alive in the ring of
workers.

Reference~\cite{fgcs} does not specify a frame return
protocol, since DIT has no result return. However, frame
return resembles stealing insofar as data (a result or loot,
respectively) are moved from one worker to another. Therefore, a
subset of the steal protocol from~\cite{fgcs} (starting after
receipt of the steal request) can be used for this purpose. The
protocol includes two checkpoints and a temporary frame saving in the
resilient store.

\section{Related Work}\label{sect:related}

There has been growing interest in task-level resilience in recent
years. Topics include task re-execution after silent
errors~\cite{ftx2,sriRaj,krishnaSC14,ftHybridTask}, techniques to
handle different failure types together~\cite{ompssResi}, and the
tracking of global data accesses of tasks~\cite{idempotent}.  While we
focus on the recovery of a dynamic task structure, Lion and
Thibault~\cite{thibault} concentrate on checkpointing the data that
are communicated between tasks.  Like us, they perform a localized
recovery, as do all previously discussed NFJ and DIT resilience
schemes and their precursors~\cite{krishna,fgcs,ijnc,cilkNOW, satin,
  mcoX10, ieeeCluster}.

Outside task-level resilience, localized recovery has been realized
for MPI programs~\cite{fenixHard, ftx1,ulfmLocal}, with the help of
Fenix~\cite{fenix} and User Level Failure Mitigation
(ULFM)~\cite{ulfm}.  Another resilient in-memory store than the IMap
was discussed in~\cite{sara}.

\section{Conclusions}\label{sect:conclusions}

This extended abstract has suggested a checkpointing algorithm for
NFJs, which supports shrinking localized recovery from one or multiple
fail-stop failures of processes. It is a variant of a previous
algorithm for DIT with only few changes that
chiefly regard the contents of checkpoints. Therefore we expect the
new algorithm to perform similarly to the original one, which causes
less than~1\% running time overhead and neglectable costs for
recovery.  Future work should experimentally investigate this
expectation.

\balance
\bibliographystyle{IEEEtran}
\bibliography{IEEEabrv,bibo}
\end{document}